\documentclass[aps,pre,twocolumn,superscriptaddress]{revtex4-1}
\PassOptionsToPackage{sort&compress}{natbib}
\usepackage{graphicx}
\usepackage{float}
\usepackage{amsmath,amssymb,amsfonts}
\usepackage{array}
\usepackage{xcolor}
\usepackage{tabularx}

\usepackage[caption=false]{subfig}

\newcommand{\intR}{\int_{\mathbb R}}

\newcommand{\avg}[1]{\langle #1 \rangle}
\newcommand{\ii}{\mathrm{i}}

\begin{document}

\title{Reduction of oscillator dynamics on complex networks to dynamics on complete graphs through virtual frequencies}

\author{Jian Gao}
\email{j.gao@rug.nl}
\affiliation{Bernoulli Institute for Mathematics, Computer Science, and Artificial Intelligence, University of Groningen, P.O.\ Box 407, 9700 AK, Groningen, The Netherlands}

\author{Konstantinos Efstathiou}
\email{k.efstathiou@dukekunshan.edu.cn}
\affiliation{Division of Natural and Applied Sciences and Zu Chongzhi Center for Mathematics and Computational Science, Duke Kunshan University, No.\ 8 Duke Avenue, Kunshan 215316, China}

\begin{abstract}
We consider the synchronization of oscillators in complex networks where there is an interplay between the oscillator dynamics and the network topology.
Through a remarkable transformation in parameter space and the introduction of virtual frequencies we show that Kuramoto oscillators on annealed networks,
with or without frequency-degree correlation, 
and Kuramoto oscillators on complete graphs with frequency-weighted coupling
can be transformed to Kuramoto oscillators on complete graphs with a re-arranged, virtual frequency distribution, and uniform coupling.
The virtual frequency distribution encodes both the natural frequency distribution (dynamics) and the degree distribution (topology).
We apply this transformation to give direct explanations to a variety of phenomena that have been observed in complex networks, such as explosive synchronization and vanishing synchronization onset.
\end{abstract}
\maketitle

\section{Introduction}

Synchronization is an important natural phenomenon, that is relevant in many processes, 
such as the flashing of fireflies \cite{Ermentrout1991},
pacemaker cells in the heart \cite{Taylor2010spontaneous},
and synchronous neural activities \cite{Bechhoefer2005feedback}.
In addition, synchronization also has practical importance in aspects of modern life, such as the functioning of power grids which is based on the synchronization of power generators \cite{Motter2013spontaneous}.
With various applications in physics, biology, and social systems,
Kuramoto-like oscillators are the most widely employed and useful analytical models for the exploration of synchronization \cite{Rodrigues2016}. 

When oscillators with different natural frequencies are connected in a complex network the interplay between the natural frequency distribution (dynamics) and the degree distribution (topology) leads to several phenomena that are not found in the standard Kuramoto model on a complete graph.
For example, recently, \emph{explosive synchronization} has been found in scale-free networks where each oscillator's natural frequency is linearly correlated with its degree \cite{Gomez2011explosive}.
Such transition process is first-order-like, discontinuous and irreversible, and is closely related to explosive percolation and cascading failures \cite{Boccaletti2016explosive}.
Explosive synchronization has also been found in oscillators on a complete graph with frequency-weighted coupling \cite{Hu2014exact}.
At the same time, oscillators on scale-free networks without frequency-degree correlation  exhibit the opposite phenomenon, that is, a continuous transition with vanishing onset \cite{Ichinomiya2004frequency}. 
In both cases the scaling exponent $\gamma$ of the scale-free networks is a critical parameter \cite{Coutinho2013kuramoto, Ichinomiya2004frequency}.
Even though these phenomena have been extensively studied \cite{Zou2014basin, Zhang2015explosive, Xu2015}, their mechanism is still unclear.

In this work, we approach the study of such systems through the self-consistent method.
For certain systems defined on complex networks or with non-uniform coupling we introduce parameter transformations that change the self-consistent equation to the one for Kuramoto systems on complete graphs.
The transformations incorporate the natural frequency distribution and degree (or coupling strength) distribution of the original system into a new distribution of quantities, which we call \emph{virtual frequencies} since they play the role of natural frequencies in the derived Kuramoto system.
The particular cases we consider include scale-free networks with or without frequency-degree correlation (where explosive synchronization and vanishing onset are found), and frequency-weighted coupling models (exhibiting explosive synchronization).
By reducing the study of Kuramoto-like oscillators on complex networks to that of Kuramoto oscillators on complete graphs we give straightforward explanations of the different dynamical phenomena that appear based on the properties of the virtual frequency distribution. 

The outline of the paper is as follows.
In Sec.~\ref{sec/self-consistent-method} we review the self-consistent method for the Kuramoto model on complete graphs.
In Sec.~\ref{sec/virtual-frequencies} we present the virtual frequency method in annealed networks, first, for networks with linear frequency-degree correlation and, second, for networks with no frequency-degree correlation.
We then apply the method to provide an alternative explanation for explosive synchronization and the vanishing onset.
In Sec.~\ref{sec/frequency-weighted-coupling} we present the virtual frequency method for networks with frequency-weighted coupling and we use it to explain explosive synchronization in this context.
We conclude in Sec.~\ref{sec/discussion} with a discussion of the limitations of the method and directions for further research.

\vfill

\section{Self-consistent method}
\label{sec/self-consistent-method}

We first review the self-consistent method as it applies to the Kuramoto model with all-to-all coupling and arbitrary (not necessarily unimodal) natural frequency distribution.
The basic idea of the Kuramoto model \cite{Kuramoto1987} to explore synchronization is to consider a group of coupled oscillators with different natural frequencies as
\begin{equation}\label{Eq.1}
  \dot{\theta_i}=\omega_i-\frac{\lambda}{N}\sum_{j=1}^{N}\sin(\theta_i-\theta_j),\, 1\leq i\leq N,
\end{equation}
where $\theta_i$ is the oscillator's phase, and $\omega_i$ is its natural frequency.
The coupling strength is given by $\lambda$, and $N$ is the size of the system.
To describe the coherent state of oscillators, the order parameter
\begin{align*}
  r \exp(\ii\phi) = \frac{1}{N} \sum_{j=1}^{N}\exp(\ii\theta_j)
\end{align*}
is introduced.
Using the order parameter, the dynamics in Eq.~\eqref{Eq.1} can be rewritten in mean field form as
\begin{equation}\label{Eq.3}
  \dot{\theta_i}=\omega_i-\lambda r\sin(\theta_i-\phi),\, 1\leq i\leq N.
\end{equation}

In this work, we are only interested in the steady states where $r(t)=r>0$ is constant and $\phi = \Omega t+\phi_0$.
In this case, analytical results on the onset of synchronization can be obtained from the analysis of each single oscillator through the self-consistent method \cite{Kuramoto1987, Strogatz2000}.
The dynamics in Eq.~\eqref{Eq.3} can be further rewritten in the frame rotating as $\Omega t+\phi_0$ and using a rescaled time $\tau = (\lambda r) t$, as
\begin{equation}\label{Eq.4}
  \dot{\theta} = b-\sin\theta,
\end{equation}
where
\begin{align*}
  b = \frac{\omega - \Omega}{\lambda r},
\end{align*}
and we have suppressed the indices of oscillators.
When $|b| \le 1$ the oscillator synchronizes with the mean field (it is locked, with phase $\theta_l$ given by $\sin\theta_l = b$ and $\cos\theta_l = \sqrt{1-b^2}$), while if $|b|>1$ it keeps running.
In the latter case, the average values of $\cos\theta$ and $\sin\theta$ are given by
\begin{align*}
  \avg{\cos\theta_r} = 0, \quad\text{and}\quad \avg{\sin\theta_r} = b \left( 1 - \sqrt{1-\frac{1}{b^2}}\right).
\end{align*}

In the continuous limit $N \to \infty$, combining steady states of these two kinds of oscillators, we obtain the self-consistent equations for the parameters $r$ and $\Omega$.
Denoting by $g_\omega(\omega)$ the distribution of natural frequencies the self-consistent equations become
\begin{equation}
  \begin{aligned}\label{Eq.5}  
  r & = \intR g_\omega(\omega) \left( \mathbf{1}_{b} \cos\theta_l 
  + (1-\mathbf{1}_{b}) \avg{\cos\theta_r} \right) \, d\omega,
  \\    
  0 & = \intR g_\omega(\omega) \left( \mathbf{1}_{b} \sin\theta_l 
  + (1-\mathbf{1}_{b}) \avg{\sin\theta_r} \right) \, d\omega,
  \end{aligned}
\end{equation}
where the indicator function $\mathbf{1}_{b}$ takes the value $1$ if $|b| \le 1$ corresponding to locked oscillators, and $0$ otherwise.
Therefore, we obtain the self-consistent equations
\begin{equation}
  \begin{aligned}\label{Eq.5b}
  \frac{1}{\lambda} & = \frac{1}{q} \intR g_\omega(\omega) \mathbf{1}_{b} \sqrt{1-b^2} \, d\omega,
  \\    
  0 & = \intR g_\omega(\omega) \left[\mathbf{1}_{b}b + (1-\mathbf{1}_{b}) b \left(1-\sqrt{1-\frac{1}{b^2}}\right)\right]  \, d\omega,
  \end{aligned}
\end{equation}
where we have divided both sides of the first self-consistent equation by $q = \lambda r$.
Further details on the self-consistent method can be found in \cite{Kuramoto1987, Strogatz2000, Acebron2005, Rodrigues2016}.
The discussion and the notation here have been adapted from \cite{Gao2018a}.

\section{Virtual frequencies in annealed networks}
\label{sec/virtual-frequencies}

In complex networks, the model for coupled oscillators reads
\begin{equation}
  \dot{\theta}_{i}
  = \omega_i + \lambda\sum_{j=1}^{N}A_{ij}\sin(\theta_{j}-\theta_{i}),\,\, i=1,\dots,N.
\end{equation}
The adjacency matrix $A_{ij}$ describes the connection of oscillators.
If there is a link between the oscillator $i$ and $j$, we have $A_{ij}=1$, and $A_{ij}=0$ otherwise.
For uncorrelated networks with randomly picked links and large order $N$ (annealed networks), the adjacency matrix can be approximated with the mean field assumption 
\begin{align*}
A_{ij} = \frac{k_ik_j}{N\avg{k}},
\end{align*}
where $k_i$ is the degree of the $i$-th node (oscillator) and $\avg{k}$ is the mean degree \cite{Ichinomiya2004frequency, Boccaletti2016explosive}.
The model now reads
\begin{equation}\label{Eq.2.1}
  \dot{\theta}_{i}
  = \omega_i + \lambda\sum_{j=1}^{N} \frac{k_ik_j}{N\avg{k}}\sin(\theta_{j}-\theta_{i}),\,\, i=1,\dots,N.
\end{equation}
A generalized order parameter (mean field) can be defined as
\begin{equation}\label{Eq.2.2}
r \exp(\ii\phi) = \sum_{j=1}^N \frac{k_j}{N\avg{k}} \exp(\ii\theta_j).
\end{equation}
Substituting the order parameter into Eq.~\eqref{Eq.2.1}, we obtain
\begin{equation*}
  \dot{\theta_i}
  = \omega_i - \lambda r k_i \sin(\theta_i-\phi),
  \,\, 1 \le i \le N,
\end{equation*}
which then reduces to the same mean field form as Eq.~\eqref{Eq.4} with  parameter
\begin{align*}
  b = \frac{\omega-\Omega}{k \lambda r}
\end{align*}
depending on both natural frequency $\omega$ and degree $k$.

\begin{figure}
  \includegraphics[width=1\linewidth]{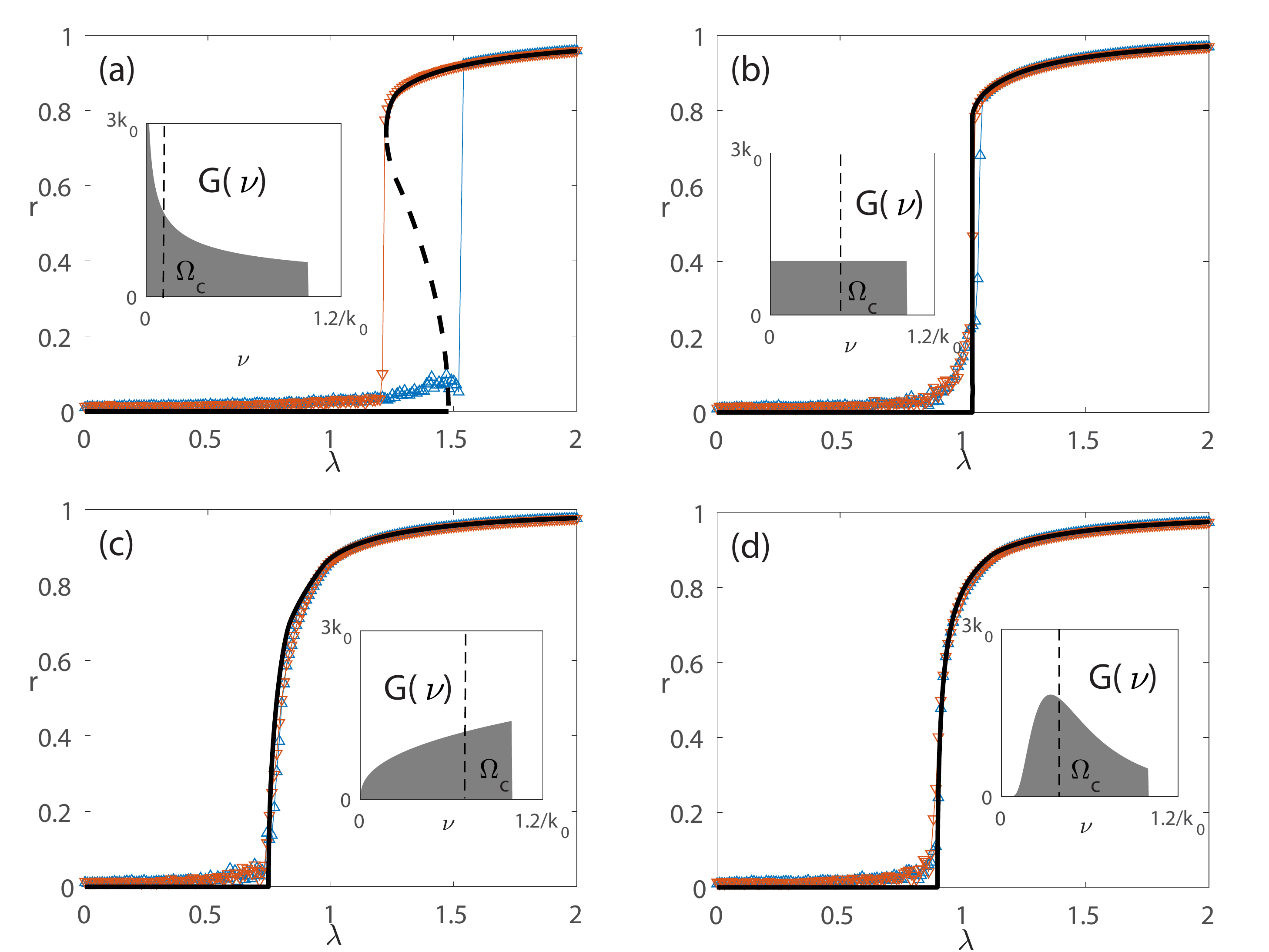}%
  \caption{Transition processes ($r$ vs $\lambda$) and re-arranged distributions $G(\nu)$ of oscillators with the frequency-degree correlation $\omega = Ak$, $A=\avg{k}^{-1}$, in annealed complex networks: (a-c) scale-free networks $g(k) \sim k^{-\gamma}$ with (a) $\gamma=2.6$, (b) $\gamma=3$, and (c) $\gamma=3.4$; (d) random network with exponential degree distribution $g(k) \sim e^{-k}$. The minimum degree is assumed $k_0=50$. Theoretical predictions (solid and dashed curves) are compared to numerical results obtained for $N=10000$ oscillators in the forward process (blue $\triangle$ with increasing $\lambda$) and backward process (red $\triangledown$ with decreasing $\lambda$).}
  \label{Fig.1}
\end{figure}

\subsection{Linear frequency-degree correlation}
\label{sec/linear-correlation}

We first consider the case where each oscillator's natural frequency $\omega$ is linearly correlated to its degree $k$ as $\omega = A k$.
The corresponding self-consistent equations, obtained by considering the continuous limit of Eq.~\eqref{Eq.2.2}, read
\begin{equation}\label{Eq.2.0}
\begin{aligned}
\frac{1}{\lambda}
& = \frac{1}{q \avg{k}}\int_0^\infty k \, g_k(k) \mathbf{1}_b \sqrt{1-b^2} \, dk,
\\
0
& = \int_0^\infty k \, g_k(k) \left[\mathbf{1}_b b + (1-\mathbf{1}_b) b \left(1-\sqrt{1-\frac{1}{b^2}}\right) \right] \, dk,
\end{aligned}
\end{equation}
where 
\begin{align*}
  b = \frac{Ak-\Omega}{kq},
\end{align*}
and $g_k(k)$ is the degree density, cf.~\cite{Coutinho2013kuramoto}.

We then define new parameters $(\Lambda, Q, W)$ and a \emph{virtual frequency} $\nu$ by
\begin{equation}\label{Eq.2.5}
\nu = \frac{1}{k}, \, W = \frac{A}{\Omega}, \, Q = -\frac{q}{\Omega}, \, \Lambda = -\frac{\lambda}{\Omega}.
\end{equation}
In terms of the new parameters and the virtual frequency we have
\[
  b = \frac{\nu-W}{Q}.
\]
Substituting the new parameters in Eq.~\eqref{Eq.2.0} leads to
\begin{equation}\label{eq/self-consistent-3}
\begin{aligned}
\frac{1}{\Lambda}
& = \frac{1}{Q} \int_0^\infty G(\nu) \mathbf{1}_b \sqrt{1-b^2} \, d\nu,
\\
0
& = \int_0^\infty G(\nu) \left[\mathbf{1}_b b + (1-\mathbf{1}_b) b \left(1-\sqrt{1-\frac{1}{b^2}}\right)\right]\,d\nu,
\end{aligned}
\end{equation}
where we have defined
\begin{equation}\label{Eq.2.6}
G(\nu) = \frac{1}{\avg{k}} \frac{1}{\nu^3} g_k\left(\frac{1}{\nu}\right).
\end{equation}
This is the same form as Eq.~\eqref{Eq.5b} which holds for complete graphs, where $(\lambda, q, \Omega, \omega)$ are replaced by $(\Lambda, Q, W, \nu)$ and the natural frequency density $g_\omega(\omega)$ is replaced by the new function $G(\nu)$.
Since $G(\nu)$ appears in Eq.~\eqref{eq/self-consistent-3} in exactly the same way as $g_\omega(\omega)$ appears in Eq.~\eqref{Eq.5b} we call the corresponding quantities $\nu$ \emph{virtual frequencies} and we call the function $G(\nu)$ \emph{virtual frequency density}.
Therefore, the self-consistent equations for the systems we consider here become the self-consistent equations for Kuramoto oscillators on complete graphs and re-arranged frequency density $G(\nu)$.

As a demonstration of the kind of understanding that can be offered by the virtual frequency method we briefly explore the phenomenon of explosive synchronization in networks with linear frequency-degree correlation.
We refer to \cite{Coutinho2013kuramoto} for a more thorough discussion of explosive synchronization in this context.

The re-arranged distribution $G(\nu)$ is determined by the degree distribution $g_k(k)$.
Depending on the divergence of quadratic mean degree $\avg{k^2}$ of $g_k(k)$, there is a clear distinction between two types of $G(\nu)$.
Consider, for example, scale-free networks with $g_k(k)\sim k^{-\gamma}$.
Then Eq.~\eqref{Eq.2.6} gives
\begin{equation}\label{Eq.2.7}
  G(\nu)= C \nu^{\gamma-3},
\end{equation}
with $\nu \in (0,1/k_0]$, where $k_0$ is the minimum degree of the network, and $C$ is the normalization factor.

For $\gamma = 3$ the distribution $G(\nu)$ is uniform with $\nu \in (0,1/k_0]$, see inset in Fig.~\ref{Fig.1}(b).
From well-known results of Kuramoto oscillators on complete graphs \cite{Strogatz2000}, the uniform distribution of natural frequencies (corresponding to $\gamma = 3$) has a hybrid synchronization transition which is abrupt and without hysteresis.
This synchronization transition is shown in Fig.~\ref{Fig.1}(b) where we compare the theoretical results obtained by solving the self-consistent equations with virtual frequencies to numerical results obtained for networks with $N=10000$ oscillators generated by the static model in \cite{Goh2001}.

For $2 < \gamma < 3$, corresponding to divergent $\avg{k^2}$, $G(\nu)$ is monotonically decreasing with $\nu$ and divergent at $\nu=0$, see inset in Fig.~\ref{Fig.1}(a).
Thus the weight of oscillators with large degrees is dramatically enlarged when $2 < \gamma < 3$.
In this case there is discontinuous transition with hysteresis (explosive synchronization) as has been earlier reported in \cite{Coutinho2013kuramoto}.
The transition for this case is shown in the comparison of theoretical and numerical results in Fig.~\ref{Fig.1}(a).

Finally, for $\gamma > 3$, corresponding to convergent $\avg{k^2}$, $G(\nu)$ is monotonically increasing and stays finite in the region $\nu \in (0, 1/k_0]$, see inset in Fig.~\ref{Fig.1}(c).
In this case the transition is continuous \cite{Coutinho2013kuramoto}, see Fig.~\ref{Fig.1}(c).
Consider now any degree distribution $g_k(k)$ which falls for large enough $k$ faster than $k^{-3}$ so that $\avg{k^2}$ is finite.
Moreover, we require that $k^3 g_k(k)$ is monotonically decreasing for large enough $k$.
It follows directly from Eq.~\eqref{Eq.2.6} that such distributions $g_k(k)$ are monotonically increasing for sufficiently small $\nu > 0$.
Consequently, for networks with several common kinds of distributions (power law, exponential, uniform, Gaussian) one gets either monotonically increasing or unimodal distributions $G(\nu)$ that give continuous transitions similarly to scale-free networks with $\gamma > 3$.
The example of the exponential distribution is shown in Fig.~\ref{Fig.1}(d). 

The continuity of the transition depends on the concavity of the distribution $G(\nu)$.
For networks with truncated distributions $g(k) \sim k^{-\gamma}$, $k_0 \le k \le k_{\mathrm{max}}$, one gets the same concavity as the original $G(\nu)$, described by $G''(\nu)$, which depends only on $\gamma$.
As a result, for finite size networks---as the one we used in numerical simulations---explosive, hybrid and continuous transitions can also be found.
This is contrary to the phenomenon of vanishing onset in scale-free networks as we discuss in the next section.

\begin{figure}
\includegraphics[width=1\linewidth]{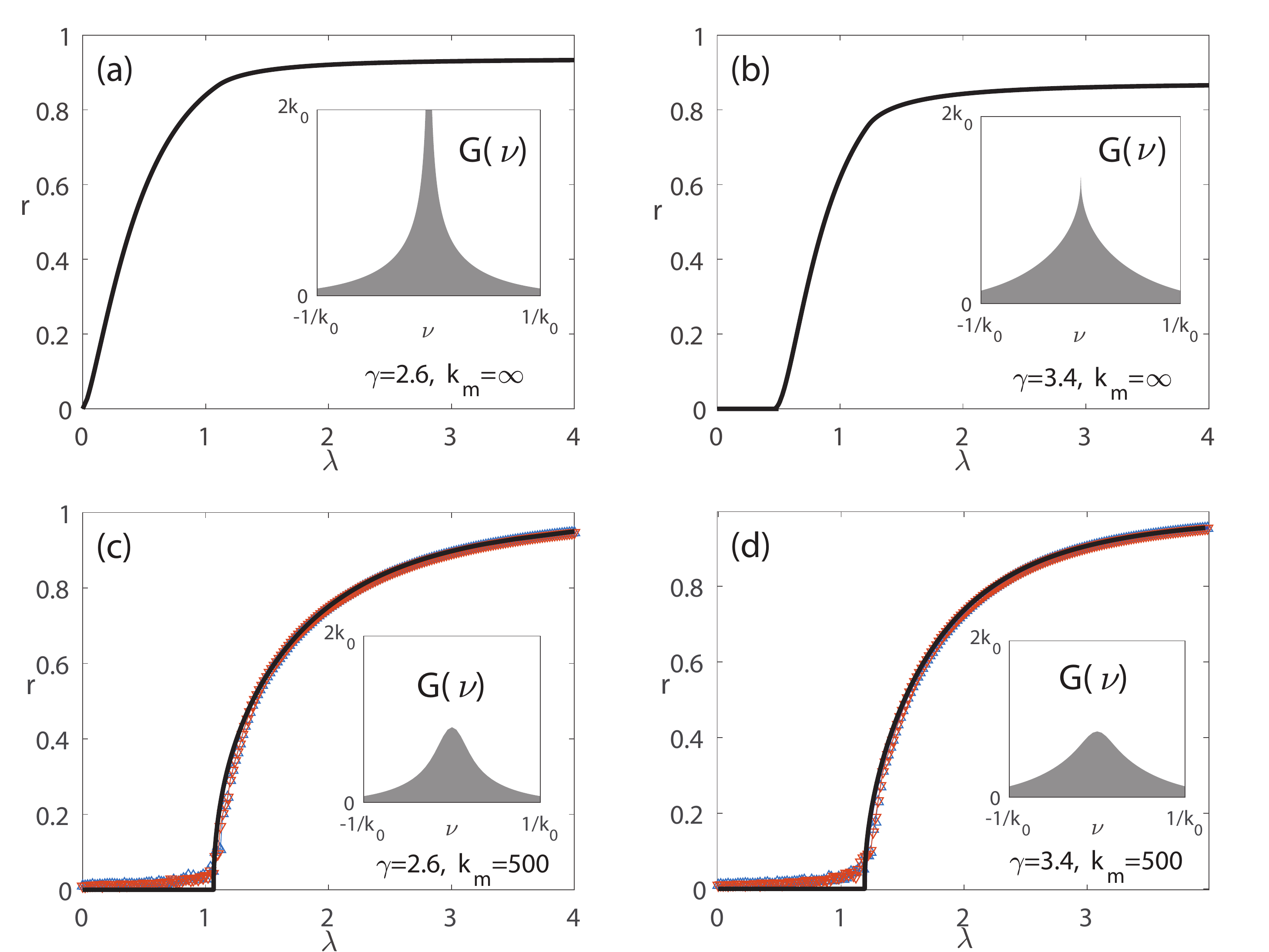}%
\caption{Transition processes ($r$ vs $\lambda$) and re-arranged distributions $G(\nu)$ of oscillators without frequency-degree correlation in annealed complex networks:  scale-free networks $g(k)\sim k^{-\gamma}$ with (a, c) $\gamma=2.6$ and (b, d) $\gamma=3.4$. The minimum degree is $k_0=50$. The maximum degree is $k_{\mathrm{m}}=\infty$ (a, b) and $k_{\mathrm{m}}=500$ (c, d). In all cases the distribution of natural frequencies is Gaussian, $g_\omega(\omega) = (1/\sqrt{2\pi\sigma^2}) \exp(-\omega^2/2\sigma^2)$ with $\sigma=\langle k\rangle$. In (c, d) the theoretically obtained curves are compared to numerical results for $N=10000$ oscillators, shown as in Fig.~\ref{Fig.1}.}
\label{Fig.2}
\end{figure}

\subsection{No frequency-degree correlation}
\label{sec/no-correlation}

Another type of system where the virtual frequency method can be applied is oscillators on a complex network where the distribution of natural frequencies $g_\omega(\omega)$ is unimodal and $g_\omega(-\omega) = g_\omega(\omega)$.
Then the dynamical equations Eq.~\eqref{Eq.2.1} imply that $\Omega = 0$ and the first self-consistent equation becomes
\begin{equation}\label{Eq.3.1}
\frac{1}{\lambda}
= \frac{1}{q} \int_{k_0}^\infty \frac{k}{\avg{k}} g_k(k)
\left(\intR g_\omega(\omega)\mathbf{1}_b\sqrt{1-b^2} d\omega\right) dk,
\end{equation}
where $k_0$ is the minimum degree of the network.
In this case, given that $b=\omega/kq$, we define the virtual frequency as 
$\nu = \omega/k$ and a corresponding virtual frequency distribution as
\begin{equation}
G(\nu) = \int_{k_0}^\infty \frac{k^2}{\avg{k}} g_k(k) g_\omega(k\nu) \, dk.
\end{equation}
With these choices, the self-consistent equation Eq.~\eqref{Eq.3.1} becomes
\[
\frac{1}{\lambda} = \frac{1}{q} \intR G(\nu) \mathbf{1}_b \sqrt{1-b^2} d\nu,
\]
where $b = \nu/q$, that is, it takes the same form as the self-consistent equation for Kuramoto oscillators on complete graphs with unimodal and symmetric (virtual) frequency density $G(\nu)$. 
Therefore, the interplay of the natural frequency density $g_\omega(\omega)$ (dynamics) and the degree density $g_k(k)$ (topology) is expressed through the re-arranged virtual frequency density $G(\nu)$.

As an example, consider the uniform distribution $g_\omega(\omega)=1/2$ with $\omega \in [-1,1]$.
For scale-free networks $g_k(k) \sim k^{-\gamma}$, with $\gamma \ne 3$, we obtain the symmetric and unimodal distribution density 
\begin{equation}
  G(\nu) = C \big(|\nu|^{\gamma-3} - k_0^{3-\gamma}\big),
\end{equation}
where $\nu \in [-1/k_0,1/k_0]$, and $C$ is the normalization constant (negative for $\gamma>3$ or positive for $2 < \gamma < 3$).
When $2< \gamma < 3$, $G(\nu)$ diverges at $\nu=0$, while for $\gamma > 3$, the distribution density remains finite.
In addition, for $\gamma = 3$ one finds $G(\nu) = C \ln(|\nu|k_0)$ for $\nu \in [-1/k_0,1/k_0]$.

The transition onset of Kuramoto oscillators with unimodal and symmetric (virtual) frequency density $G(\nu)$ is determined by $\lambda_c = 2/\pi G(0)$.
Therefore, the divergence of $G(\nu)$ at $\nu = 0$ for $2 < \gamma \le 3$ results to $\lambda_c = 0$, that is, vanishing onset.
The transition processes and corresponding virtual frequency distributions are shown in Fig.~\ref{Fig.2}(a-b) for Gaussian natural frequency distributions and scale-free networks.

The previous discussion can be extended to other types of networks.
Networks can be divided into two categories depending on the divergence of the quadratic mean degree $\avg{k^2}$.
If and only if $\avg{k^2}$ is convergent (e.g., for exponential degree distributions), the virtual frequency distribution $G(\nu)$ remains finite at $\nu = 0$, similar to the case $\gamma > 3$, and thus we do not have vanishing onset.
This result was previously obtained in \cite{Ichinomiya2004frequency}.

Since the vanishing onset depends on the convergence of $\avg{k^2}$, it is sensitive to the tail of the distribution.
For example, for \emph{broad-scale networks} with truncated distributions $g_k(k) \sim k^{-\gamma}$, $k_0 \le k \le k_{\max}$, the corresponding $\avg{k^2}$ is finite, and thus the virtual frequency distribution $G(\nu)$ is also finite at $\nu = 0$, as shown in Fig.~\ref{Fig.2}(c-d). 
Note that any finite system has a maximum degree $k_{\max}$.
Hence the vanishing onset can only be observed for systems with $N \to \infty$.

\begin{figure}
	\includegraphics[width=1\linewidth]{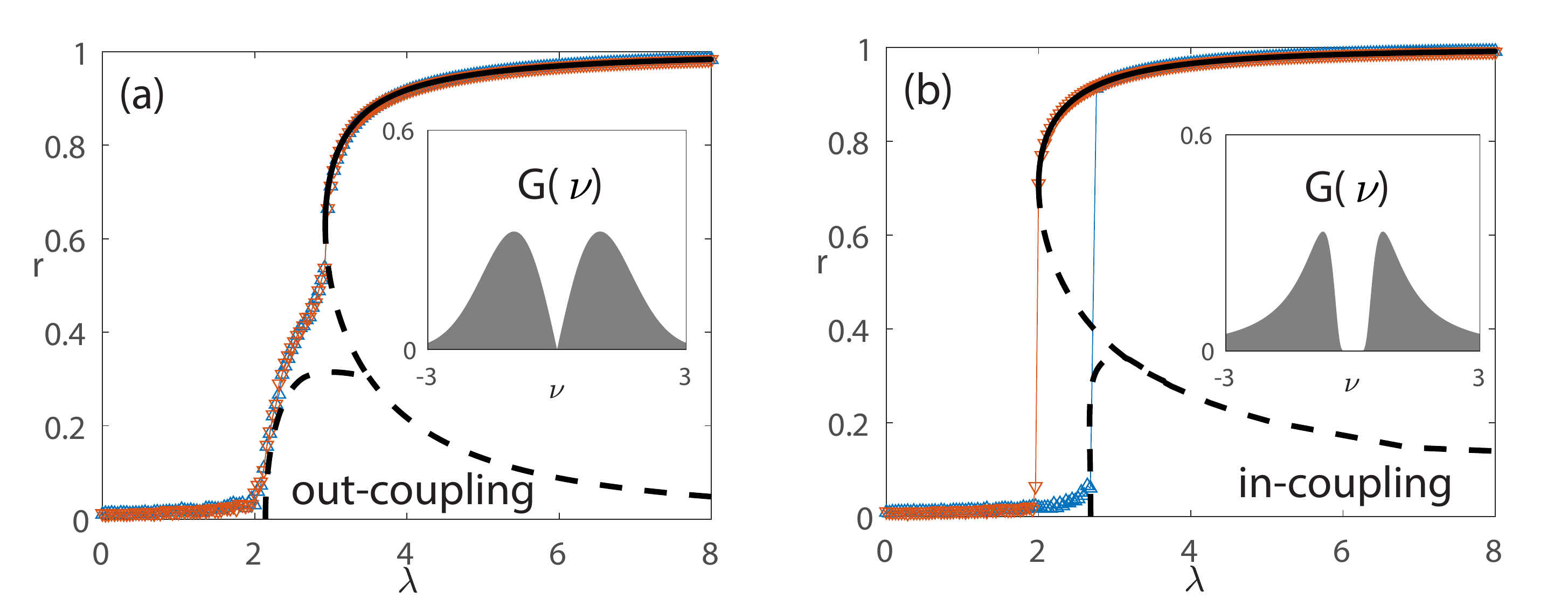}%
  \caption{Transition processes ($r$ vs $\lambda$) and re-arranged distributions $G(\nu)$ of frequency-weighted-coupling oscillators: 
  (a) out-coupling model and (b) in-coupling model. 
  The natural frequencies density is Gaussian, $g_\omega(\omega) = (1/\sqrt{2\pi}) \exp(-\omega^2/2)$. 
  The theoretically obtained curves are compared to numerical results for $N=10000$ oscillators, shown as in Fig.~\ref{Fig.1}.}
	\label{Fig.3}
\end{figure}

\section{Networks with frequency-weighted coupling}
\label{sec/frequency-weighted-coupling}

Except for the model with frequency-degree correlation in scale-free networks, another model that exhibits explosive synchronization, is the Kuramoto model with absolute frequency-weighted coupling \cite{Hu2014exact,Xu2016synchronization}. It is defined on complete graphs as
\begin{equation}\label{Eq.4.1}
\dot{\theta_i}=\omega_i-\frac{\lambda}{N}\sum_{j=1}^{N}F_{ij}\sin(\theta_i-\theta_j),\, 1\leq i\leq N,
\end{equation}
where $F_{ij}=|\omega_i|$ (in-coupling model) or $F_{ij}=|\omega_j|/\langle |\omega|\rangle$ (out-coupling model), mimicking the frequency-degree correlation \cite{Bi2016coexistence,Xu2018origin}.
The frequency-weighted coupling model typically shows explosive synchronization (and also oscillatory states, such as standing waves and Bellerophon states) \cite{Xu2018origin, Bi2016coexistence}.

For the out-coupling model, an order parameter is defined as
\[
r \exp(\ii \phi) = \sum_{j=1}^N \frac{|\omega_j|}{N \avg{|\omega|}} \exp(\ii\theta_j).
\]
Encoding the frequency-weighted coupling, the self-consistent equation can be rewritten, using the virtual frequencies $\nu = \omega$, in standard form with re-arranged distribution 
\[
  G(\nu)= \frac{|\nu| g_\omega(\nu)}{\avg{|\omega|}}.
\] 
For any normalized distributions $g_\omega(\omega)$, $G(\nu) \to 0$ as either $\nu \to 0$ or $\nu \to \pm\infty$.
Thus for any unimodal symmetric distribution $g_\omega(\omega)$ the re-arranged distribution $G(\nu)$ is bimodal and symmetric,
see Fig.~\ref{Fig.3}(a).

For the in-coupling model, the case becomes more complicated.
For the steady-state solution with $\Omega=0$, we have $b=\omega/|\omega|q$ and thus we define the virtual frequency $\nu=\operatorname{sign}(\omega)$, which is naturally bimodal.
For $\Omega \ne 0$,  we define the virtual frequency through the transformation
\begin{equation}\label{Eq.4.3}
\nu = \frac{1}{\omega}, \, W = \frac{1}{\Omega}, \, Q = \frac{q}{\Omega}\operatorname{sign}(\omega), \, \Lambda = \frac{\lambda}{\Omega}\operatorname{sign}(\omega),
\end{equation}
with density
\begin{equation}\label{Eq.4.4}
  G(\nu)=\frac{1}{\nu^2}g_\omega\left(\frac{1}{\nu}\right).
\end{equation}
The latter is bimodal and symmetric when $g_\omega(\omega)$ is unimodal and symmetric, see Fig.~\ref{Fig.3}(b).
Note, that in this case the coupling strength $\Lambda$ can be either positive or negative, unlike the standard Kuramoto model. 

For coupled oscillators, bimodal frequency distributions and the coexistence of the positive and negative coupling strength contribute to abrupt transitions and oscillatory states (standing wave, $\pi$ state) \cite{Hong2011kuramoto, Martens2009exact}.
The frequency-weighted coupling model, especially the in-coupling one, includes these two factors and hence one can anticipate its explosive synchronization and the existence of oscillatory (Bellerophon) states \cite{Bi2016coexistence}.
The details of this relation can be analyzed in a more general framework, where the self-consistent method is related to non-steady states \cite{Gao(inpreparation)}.

\section{Discussion}
\label{sec/discussion}

We have shown that with appropriate transformations, certain oscillator systems on complex networks are transformed to the standard Kuramoto model on complete graphs with a re-arranged virtual frequency distribution.
Such distributions combine the effect of topology, dynamics, and their correlation, leading to a deeper intuitive understanding of the onset of synchronization.
Our method can be generalized to more complicated cases, such as the partial degree-frequency correlation \cite{Pinto2015explosive} and the degree correlations \cite{Sendina2015effects, Restrepo2014mean}.
Including such systems, we can obtain a more general framework of Kuramoto-like synchronization, whereas the models studied in this work are the linear cases \cite{Gao(inpreparation)}.

However, there are also situations where the method of virtual frequencies cannot be applied without modifications.
In particular, our analysis is based on the self-consistent method and is assuming either complete graphs or annealed complex networks.
We note that annealed complex networks approximate random complex networks with a large mean-degree $\avg{k} \gg 1$ \cite{Ichinomiya2005path, Sonnenschein2012onset, Peron2012determination} and therefore the method may not work equally well for sparse networks.

Another system where the virtual frequencies method cannot be applied is the Kuramoto-Sakaguchi model. Even though the Kuramoto-Sakaguchi model can be studied through the self-consistent method, the effect of phase shifts cannot be combined into the virtual frequencies snf alternative approaches are necessary.

\begin{acknowledgments}
J.G. is supported by a China Scholarship Council (CSC) scholarship.
\end{acknowledgments}

\bibliographystyle{apsrev4-2}
\bibliography{papersnew.bib}

\end{document}